# DANLU TONGDU TABLETS TREAT LUMBAR SPINAL STENOSIS THROUGH REDUCING REACTIVE OXYGEN SPECIES AND APOPTOSIS BY REGULATING CDK2/CDK4/CDKN1A EXPRESSION

Xue Bai[1,a], Ayesha T.Tahir[2,b], Zhengheng Yu[c,3], Wenbo Cheng [4,d,*], Bo Zhang [5,6,e,*], Jun Kang[1,f,*]

[1] School of life sciences, Tianjin University, No.92 Weijin Road, Nankai District, Tianjin, 300072, China

[2] Department of Biosciences, COMSATS University Islamabad, Park Road, 45550, Islamabad, Pakistan

[3] Breast Disease Center, Peking University First Hospital, Beijing, China

[4] Tianjin Key Laboratory of Medical Mass Spectrometry for Accurate Diagnosis, No.23 Wujing Road, Dongli District, Tianjin, 300399, China

[5] Institute for TCM-X, MOE Key Laboratory of Bioinformatics, Bioinformatics Division, BNRist, Department of Automation, Tsinghua University, Beijing 100084, China.

[6] Beijing Intelligent Medicine and Network Pharmacology Co., Ltd, Beijng 100020, China

[a]e-mail: baixue1219@tju.edu.cn
[b]e-mail: ayesha.tahir@comsats.edu.pk
[c]e-mail: yuzhengheng1997@163.com
[d]e-mail: chengwb@sibet.ac.cn
[e]e-mail: zhangbo.2007@tsinghua.org.cn
[f]e-mail: jun.kang@tju.edu.cn

**\* Correspondence**
\*Corresponding authors,
e-mails: jun.kang@tju.edu.cn; zhangbo.2007@tsinghua.org.cn; chengwb@sibet.ac.cn

**Abstract**—Lumbar spinal stenosis (LSS) is caused by the compression of the nerve root or cauda equina nerve by stenosis of the lumbar spinal canal or intervertebral foramen, and is manifested as chronic low back and leg pain. Danlu Tongdu (DLTD) tablets can relieve chronic pain caused by LSS, but the molecular mechanism remains largely unknown. In this study, the potential molecular mechanism of DLTD tablets in the treatment of LSS was firstly predicted by network pharmacology method. Results showed that DLTD functions in regulating anti-oxidative, apoptosis, and inflammation signaling pathways. Furthermore, the flow cytometry results showed that DLTD tablets efficiently reduced ROS content and inhibited rat neural stem cell apoptosis induced by hydrogen peroxide. DLTD also inhibited the mitochondrial membrane potential damage induced by hydrogen peroxide. Elisa analysis showed that DLTD induced cell cycle related protein, CDK2 and CDK4 and reduced CDKN1A protein expression level. Taken together, our study provided new insights of DLTD in treating LSS through reducing ROS content, decreasing apoptosis by inhibiting CDKN1A and promoting CDK2 and CDK4 expression levels.

# INTRODUCTION

Lumbar spinal stenosis (LSS) causes the shortening of the diameter lines of the spinal canal, and compression of the dural sac, spinal cord, or nerve root, resulting in neurological dysfunction [1]. LSS mostly occurs in middle-aged and elderly people over 50 years old, bringing heavy burden to patients [2].

As a kind of non-surgical treatment, Traditional Chinese Medicine (TCM) plays an increasingly important role in the treatment of LSS. According to the TCM theory, LSS is caused by liver and kidney dysfunction. Evil winds cold dampness, Qi stagnation and blood stasis, resulting in pain in the waist and legs. Therefore, TCM which could promote blood circulation, remove blood stasis, and nourish the liver and kidney, is considered an ideal drug for the treatment of LSS [3,4]. Danlu Tongdu (DLTD) tablets are a kind of TCM preparation. It clinically relieves chronic lumbago and leg pain caused by LSS. The main components of DLTD are *Salvia miltiorrhiza* Bunge (DanShen), *Astragalus membranaceus (Fisch.)* Bunge (HuangQi), *Corydalis* *yanhusuo* (YanHuSuo), *Eucommia ulmoides Oliver* (DuZhong) and *Cervus nippon Temminck* (Lu JiaoJiao). Among them, DanShen is considered as the king drug that activates blood and removes stasis, dredges menstruation and relieves pain. HuangQi, YanHuSuo and Lu JiaoJiao are official drugs, which have the effects of warm and nourish Qi and Blood, benefiting the liver and kidney. DuZhong could tonify the liver and kidney [5]. However, the underlying molecular mechanism of DLTD in treating LSS disease is still unclear.

Oxidative stress is caused by an imbalance in the homeostasis of pro-oxidants and antioxidants, leading to the production of high levels of reactive oxygen species (ROS) [6]. It has been reported that the increased production of ROS and free radicals in cellular redox levels is related to degenerative diseases and aging. Studies have shown that oxidative stress plays a pivotal role in the pathogenesis of hypertrophic ligamentum flavum in LSS [7,8]. In addition, LSS causes cauda equina nerve compression and makes patients tingling. Cauda equina compression induces apoptosis, characterized by the increased expression of p53, Bax, and Bad, and the decreased expression of Bcl-2 [9,10]. Furthermore, the production of ROS is also closely related to mitochondrial apoptosis, a high level of ROS could facilitate $Ca^{2+}$ and change the expression of apoptosis related proteins, leading to cytochrome *c* releasing from mitochondria. This process will activate caspase protein activity, ultimately leading to cell apoptosis [11-13].

Cyclin dependent kinase inhibitor 1A (CDKN1A) encodes an effective cyclin-dependent kinase inhibitor. It binds with cyclin-dependent kinase (CDK) alone or binds to the cyclin-CDK complex, interfering with the cell cycle [14]. CDK2 and CDK4 mainly act on the G1 and S phases of the cell cycle, initiate DNA replication and induce mitosis, thus promoting cell proliferation [15]. The protein encoded by CDKN1A could bind to and inhibit the activity of cyclin-CDK2 or cyclin-CDK4 complexes, respectively, thus blocking-up cell cycle [16]. However, CDKN1A could be specifically cleaved by Caspase-3, resulting in significant activation of CDK2 [17,18]. It has been shown that down regulating the expression of CDKN1A can promote G1-S phase transition, and inhibit cell apoptosis [19].

TCM network pharmacology, which is firstly proposed by Li [20,21], provide a novel systematic way to understanding TCM and is widely used in exploring mechanism underlying herbal formula [22]. In this study, we predicted the DLTD-targets-LSS network through the network pharmacology method. The effects of DLTD on reducing ROS content, cell apoptosis level, and mitochondrial membrane potential damage induced by hydrogen peroxide were analyzed. The expression level of three cell cycle related proteins, CDKN1A, CDK2, and CDK4 was also detected to verify the function of DLTD. The overall experimental process and the molecular mechanism of Danlu Tongdu are shown in **Fig. 1A, 1B**.

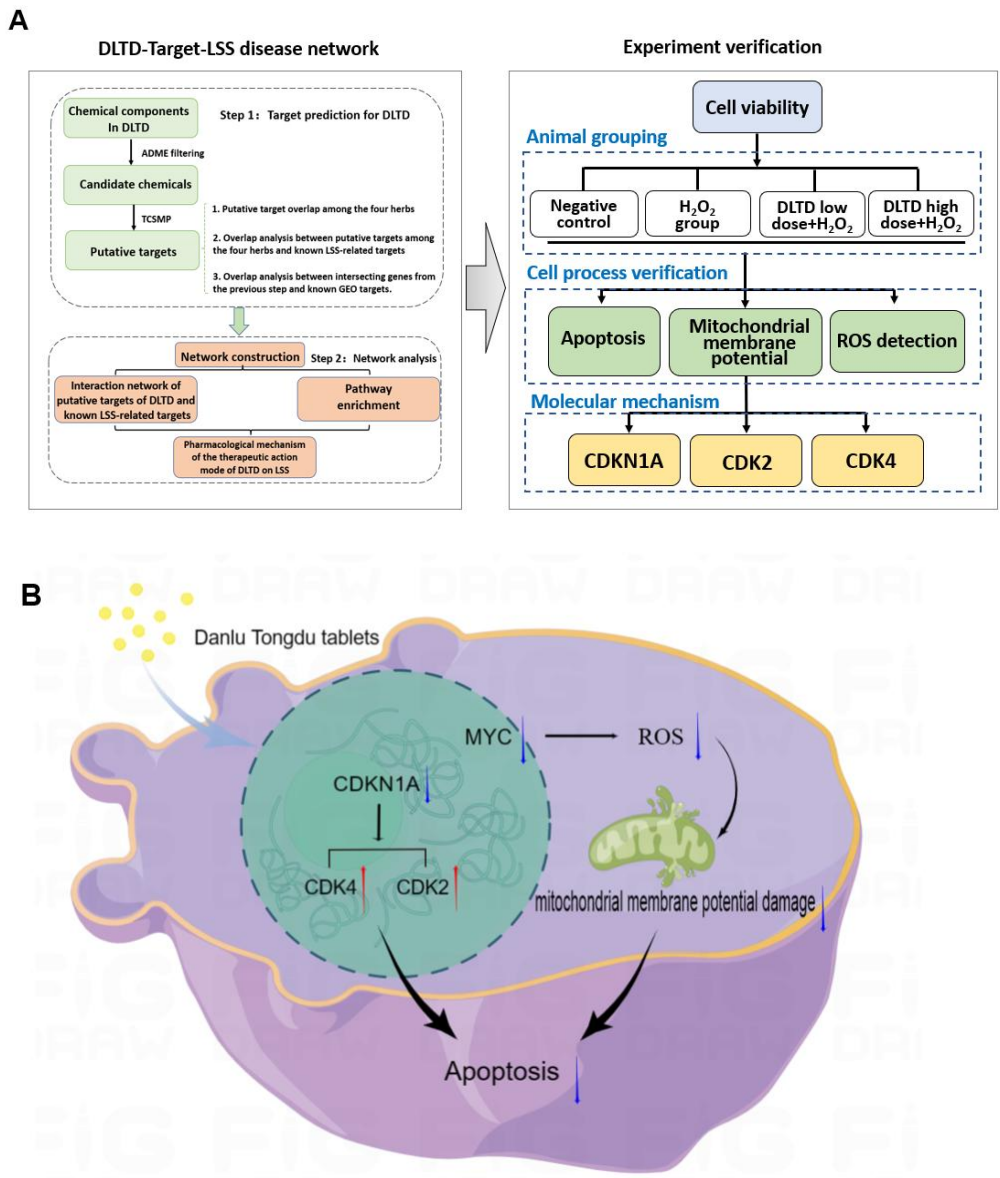

**Fig. 1.** A. The workflow of the approach for uncovering the pharmacological mechanisms of DLTD actions on LSS. B. The schematic diagram of the pharmacological mechanism of DLTD actions on LSS.

## MATERIALS AND METHODS

**Network Pharmacology analysis.** The ingredients in DLTD were searched and collected from TCMSP. Based on the collection of ingredients collected above, we used oral bioavailability (OB) ⩾ 30% and drug-likeness (DL) ⩾ 0.18 as the screening to select the active ingredients.

The target proteins of the four herbs were predicted by the DrugBank database in TCMSP. These target names were calibrated to the standardized name using the UniProt database. The target genes of LSS were searched and collected from Genecards database. Potential target genes of DLTD therapy for LSS were acquired through the Veeny intersection.

Gene Ontology (GO) and Kyoto Encyclopedia of Genes and Genomes (KEGG) were used to perform pathway enrichment analysis.

**Preparation of Danlu Tongdu lyophilized powder.** DLTD tablets were purchased from Henan Lingrui Pharmaceutical Co., Ltd., with the national drug approval No. z20050085. Take 5 pieces of DLTD tablets (3 g in total) and grind them into powder. Add 3 mL sterile water to dissolve them to obtain the suspension. After ultrasonic extraction at 800 W for 30 minutes, filter them twice with 0.45 μm and 0.22 μm filter membranes to obtain the aqueous solution of DLTD tablets. Then add freeze-drying protective agent, 10% mannitol, to the aqueous solution, mix well and freeze-drying to obtain DLTD lyophilized powder. The lyophilized powder was then dissolved in the cell culture medium to prepare a mother liquor with a concentration of 500 mg/mL for cell experiments.

**Cell culture.** Rat neural stem cell line (Procell, CP-R139) was purchased from Procell Life Science & Technology Co., Ltd. Rat neural stem cells were cultured in the special growth medium (Procell, CM-R139). Rat neural stem cells were cultured under normal conditions (5% $CO_2$, 21% $O_2$, and 74% $N_2$) in a humidified incubator at 37°C. The passage step was performed once a week using 0.25% Trypsin-EDTA (NCM Biotech, Suzhou, China) and stored in serum-free cell freezing medium CELLSAVING (NCM Biotech, Suzhou, China) under −80°C.

**Cell viability.** Cells were seeded in a 96-well plate and grew to be approximately 60% confluent. Then seeded cells were added with 0 mg/mL, 10 mg/mL, 20 mg/mL, 30 mg/mL, 40 mg/mL, 50 mg/ml DLTD and 500 μM $H_2O_2$ individually. After 8 h culture, the results were determined by cell counting kit-8 (CCK-8, Solarbio, China). All wells were treated with 10 μL CCK-8 reagent at 37 °C for 2 h. At last, the absorbance at 450 nm was measured by EnSpire Multilabel Reader.

**Cell apoptosis measurement.** The rate of cellular apoptosis was measured by the Annexin V-FITC Apoptosis Detection Kit (BD Bioscience, USA). Cells were plated and incubated for 24 h, prior to being treated with DLTD (5 mg/ml and 10 mg/ml, 24 h) and $H_2O_2$ (500 μM, 12 h). Then cells were harvested, washed with PBS, and followed the manufacturer's protocols for the detection of apoptosis by using. The distribution of cell populations in different quadrants was detected using the flow cytometer (FACSCalibur, BD, USA).

**Mitochondrial membrane potential damage detection.** Mitochondrial membrane potential damage was detected with the Mitochondrial Membrane Potential Kit (JC-10 Assay, Solarbio). Cells were plated and incubated for 24 h, prior to being treated with DLTD (5 mg/ml and 10 mg/ml, 24 h) and $H_2O_2$ (500 μM, 12 h). Then cells were harvested, washed with PBS, and followed the manufacturer's protocol.

**ROS content detection.** Cells were treated with DLTD (5mg/ml and 10mg/ml, 24h) and $H_2O_2$ (500uM, 12h), and generated ROS level was measured with 2,7-Dichlorodihydrofluorescein diacetate (DCFH-DA, Solarbio) by laser scanning confocal microscopy (LSCM).

**Elisa assay.** The Rat CDK2 ELISA kit was purchased from BIOESN (BES5216K, China), Rat CDK4 ELISA kit was purchased from RenJieBio (RJ23041,

China) and Rat CDKN1A ELISA kit was purchased from Ybscience (YB-CDKN1A-Ra, China). All experimental operations were according to the instructions.

**Statistical analysis.** Experiments were performed with three biological repetitions. Data are shown as mean ± SD. Statistical analysis was performed with *t*-test of GraphPad Prism.

## RESULTS AND DISCUSSION

**Construction of DLTD-Targets-LSS disease network.** To determine the active ingredients in DLTD, we searched its components in TCMSP database. In total, 162 active ingredients in DLTD were screened, including 65 in DanShen, 20 in HuangQi, 49 in YanHuSuo, and 28 in DuZhong (Supplemental Table S1). Lu JiaoJiao was not found in TCMSP database. Lu JiaoJiao is a solid glue made by decocting and concentrating of ossified antlers or antlers that fall off in the spring of the second year after sawmilling from *Cervus elaphus* Linnaeus or *Ceruus nippon* Temminck [23]. It was reported that Lu Jiaojiao contains animal proteins, various amino acids, peptides, hormones and polysaccharide [24]. The protein content of Lu JiaoJiao is 82.49% and the total amino acid content is 34.40%. So far, 19 kinds of amino acids, including 8 kinds of essential amino acids, have been isolated and identified from Lu JiaoJiao. Among them, glycine, proline, hydroxyproline, and alanine are the four components with the largest content, and they are also the main components of collagen [25].

**Supplemental Table S1.** The 162 active ingredients of DLTD found in TCMSP database.

The target proteins of DLTD ingredients were then predicted by the DrugBank database. Among the 162 active ingredients in DLTD, 151 ingredients (including 59 in DanShen, 17 in HuangQi, 49 in YanHuSuo, and 26 in DuZhong) were found in DrugBank database (Supplemental Table S2). We found that there were 112 target proteins of active components in DanShen, 172 in HuangQi, 172 in YanHuSuo, and 169 in DuZhong. The predicted protein targets of the four herbs were then analyzed by Venny diagram, and 78 intersecting target proteins were found in all four herbs (Fig. 2A, Supplemental Table S3).

**Supplemental Table S2.** The 151 active ingredients of DLTD found in DrugBank database.

**Supplemental Table S3.** The 78 intersecting target proteins of active components in DLTD tablets.

To find out the LSS disease related genes, we searched Genecards database with the key word LSS. 1,924 LSS target genes were collected (Supplemental Table S4). To further confirm the LSS related genes, we searched the LSS related transcriptome data from the NCBI GEO database. The sequencing data No. GSE113212 was related to LSS. We then analyzed the data with |LogFC| > 1 as the screening criteria. 3837 differentially expressed genes were obtained (Supplemental Table S5). Finally, 23 intersecting genes from DLTD-targets, LSS-targets from Genecards database, and LSS-targets from transcriptome were collected (Fig. 2B), including nitric oxide synthase 2 (NOS2), cytochrome P450 family 1 subfamily A member 1 (CYP1A1), tumor necrosis factor (TNF), secreted phosphoprotein 1 (SPP1), MYC proto-oncogene (MYC), matrix metallopeptidase 1(MMP1), heme oxygenase 1 (HMOX1), interleukin 10 (IL10), MDM2 proto-oncogene (MDM2), calcitonin receptor (CALCR), cyclin dependent kinase inhibitor 1A (CDKN1A), interleukin 1 beta (IL1B), insulin like growth factor binding protein 3 (IGFBP3), prostaglandin E receptor 3 (PTGER3), gap junction protein alpha 1 (GJA1), peroxisome proliferator activated receptor gamma (PPARG), C-X-C motif chemokine ligand 8 (CXCL8), myeloperoxidase (MPO), interleukin 6 (IL6), Fos proto-oncogene (FOS), hypoxia inducible factor 1 subunit alpha (HIF1A), prostaglandin-endoperoxide synthase 1 (PTGS1), and matrix metallopeptidase 9 (MMP9).

**Supplemental Table S4.** The 1924 target genes of LSS disease.

**Supplemental Table S5.** The 3837 differentially expressed genes in LSS from transcriptome analysis.

To explore the therapeutic mechanism of potential targets of DLTD, DAVID database was used to conduct Gene Ontology enrichment analysis of the 23 genes. A total of 3 main biological functions were obtained from GO enrichment analysis, including biological process, cellular component, and molecular function (Fig. 2C). The top 10 enriched GO analysis for biological process, including: inflammatory response, response to drugs, negative regulation of apoptotic process, and aging et al. (Supplemental Table S6). Furthermore, a total of 15 signaling pathways were obtained from the KEGG pathway enrichment analysis, including hsa05323:Rheumatoid arthritis，hsa04620:Toll-like receptor signaling pathway，hsa04066:HIF-1 signaling pathway，hsa04668:TNF signaling pathway，hsa04060:Cytokine-cytokine receptor interaction，hsa04151:PI3K-Akt signaling pathway，hsa04621:NOD-like receptor signaling pathway，hsa04068:FoxO signaling pathway，hsa04010:MAPK signaling pathway，hsa04115:p53 signaling pathway，hsa04064:NF-kappa B signaling pathway，hsa04660:T cell receptor signaling pathway，hsa04919:Thyroid hormone signaling pathway，hsa04110:Cell cycle，hsa04630:Jak-STAT signaling pathway (Fig. 2D, Supplemental Table S7). The compound-target network of candidate compounds in four herbs in DLTD was shown in Fig. 2E. And the total network of DLTD-Target-LSS disease was shown in Fig. 2F.

**Supplemental Table S6.** GO analysis of the intersecting genes of DLTD and LSS.

**Supplemental Table S7.** KEGG analysis of the intersecting genes of DLTD and LSS.

**Fig. 2.** DLTD-Target-LSS disease network. A. The Venn diagrams showing the putative target overlap among the four herbs. B. The Venny results of the potential target genes of DLTD therapy for LSS. C. The GO analysis for biological function of potential target genes of DLTD in LSS. D. The enriched KEGG analysis for the signaling pathway of potential target genes of DLTD in LSS. E. The compound-target network of candidate compounds in four herbs in the DLTD. Green circle: candidate compound; Red inverted triangle: herb; Purple diamond: target. F. The total network of DLTD-Target-LSS disease.

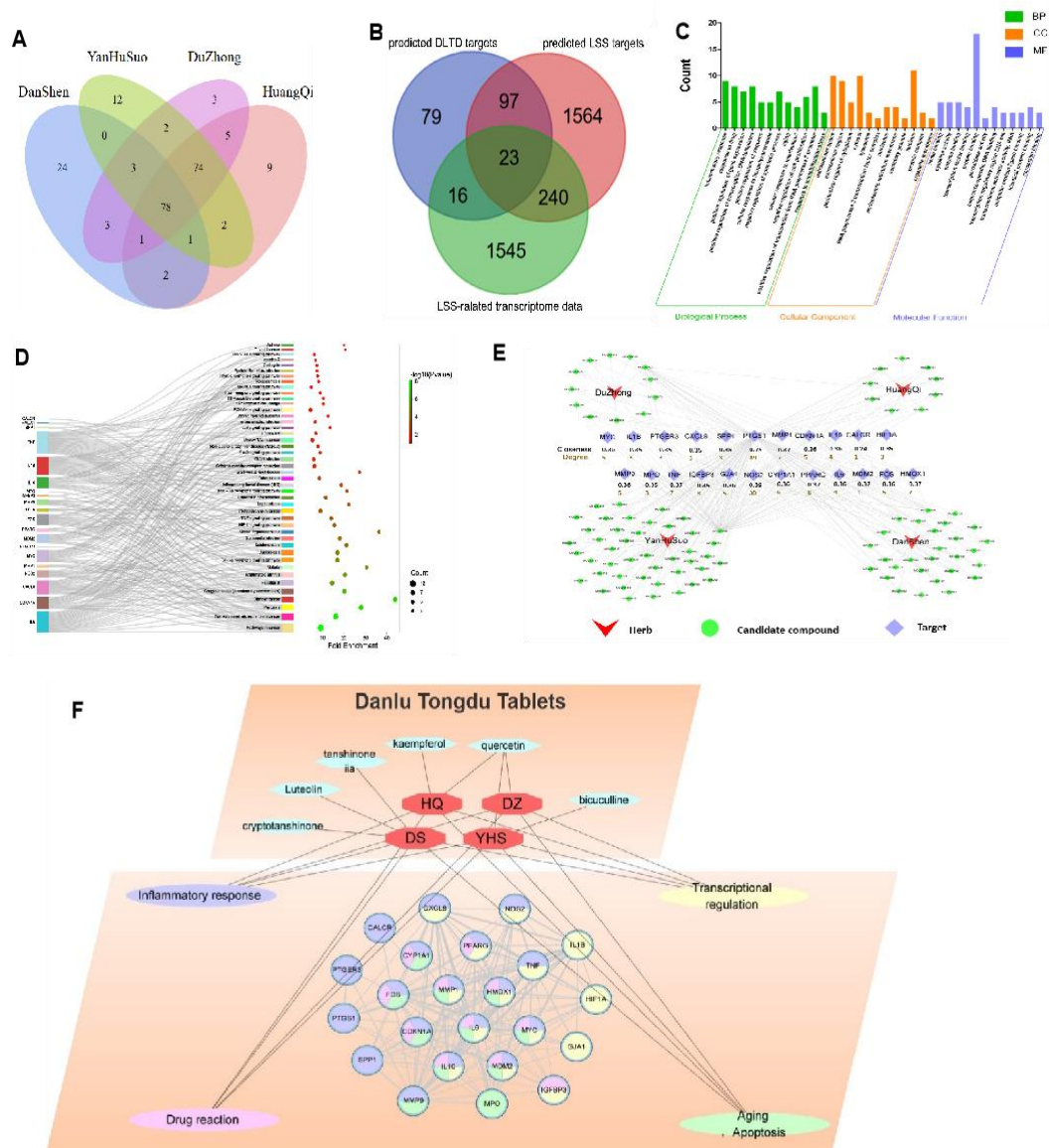

**Appropriate DLTD concentration detection.** To confirm the proper DLTD

concertation in cell experiments, rat neural stem cell viability was detected with CCK8 method. The results showed that the cell viability was ~100% with 5 mg/ml or 10 mg/ml DLTD treated (Fig. 3A). Therefore, 5mg/ml or 10mg/ml DLTD were chosen as the low dose group or the high dose group in the following experiment, respectively.

**DLTD reduced the ROS content.** Studies have shown that ROS may be one of the causes of LSS [8]. We wondered whether DLTD could reduce the ROS level in cells. Cells were firstly treated with hydrogen peroxide to cause oxidative damage, then treated with low- or high-dose of DLTD for 24 h. The results showed that the ROS content in the $H_2O_2$ group was ~38%. With low-dose DLTD treatment, the ROS content decreased to ~11%. After the high-dose DLTD treatment, the ROS content was ~9% in cells (Fig. 3B and 3C). This revealed that DLTD could significantly reduce the ROS level in cells, and in a dose dependent manner.

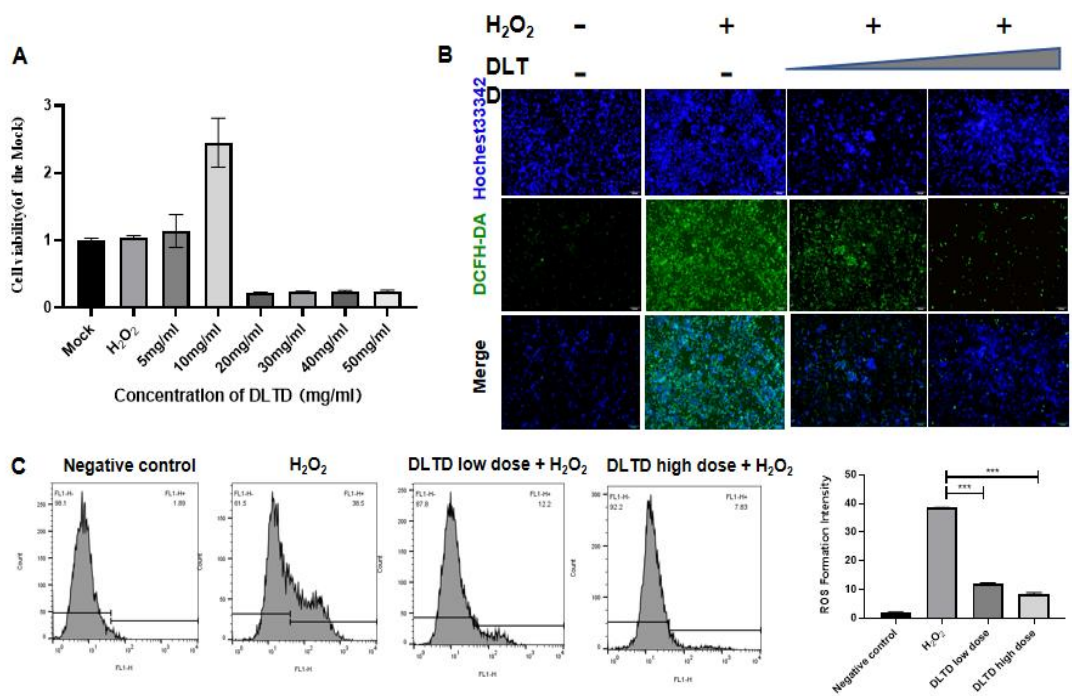

**Fig. 3.** DLTD reduce the content of ROS in cells. A. Appropriate DLTD concentration detection. Cells were treated with 0, 5, 10, 20, 30, 40, 50 mg/ml DLTD

for 24 h, respectively. Then the cell viability was analyzed by CCK8. B. Confocal fluorescence images of live cells (blue) and ROS (green) by Hochest33342 and DCFH-DA double staining after different treatments with DLTD and $H_2O_2$. C. Flow cytometry analysis the content of ROS in rat neural stem cell cells, and the fluorescence area (using the same threshold). Data are presented as the mean ± s.d. n = 3. *$P < 0.05$, **$P < 0.01$, ***$P < 0.001$.

**DLTD inhibited cell apoptosis induced by hydrogen peroxide.** Previous studies have shown that cauda equina compression caused by LSS could induce apoptosis in rats. In the network pharmacology analysis, we also found that DLTD targets the negative regulation of apoptosis pathway genes, indicating that DLTD may inhibit cell apoptosis. To investigate the function of DLTD on inhibiting apoptosis, the rat neural stem cell cells were firstly treated with 500 μM $H_2O_2$ to induce apoptosis. Then DLTD was added to the apoptotic cells. Then cells were detected by flow cytometry. The results showed that both low- and high-dose of DLTD treatment could inhibit the apoptosis caused by $H_2O_2$. The low dose group recovered 24% of the apoptosis caused by $H_2O_2$, while the DLTD high dose group recovered 87% of the apoptosis caused by $H_2O_2$ (Fig. 4).

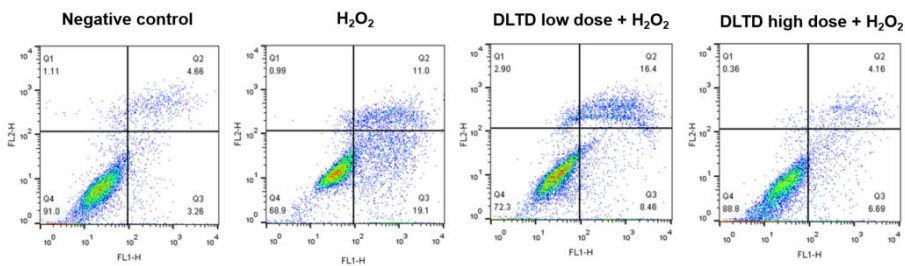

**Fig. 4.** DLTD inhibit the apoptosis of rat neural stem cell cells induced by hydrogen peroxide. A. Cell viability caused by different treatments, analyzed by CCK8, difference analysis of each group was performed with the control group. B. Flow cytometry analysis of rat neural stem cell cells apoptosis FITC/PI double staining after different treatments with DLTD and $H_2O_2$.

**DLTD inhibits the mitochondrial membrane potential damage induced by hydrogen peroxide**. Evidence showed that the process of apoptosis is closely related to mitochondria damage. It was considered that mitochondria are the executor of apoptosis [26]. Mitochondrial membrane potential (MMP) is a marker that reflects the function of the mitochondria, and its complete loss only occurs in cells to be apoptotic. To study the effect of DLTD on regulating mitochondrial membrane potential damage, the rat neural stem cell cells treated with DLTD and $H_2O_2$ were stained with JC-10 and detected by flow cytometry. The results of flow cytometry showed that the DLTD low dose group recovered 45% of the mitochondrial membrane potential damage caused by $H_2O_2$. And the DLTD high dose group recovered 55% of the mitochondrial membrane potential damage caused by $H_2O_2$ (Fig. 5A and 5B). This result indicated that DLTD may inhibit cell apoptosis by alleviating mitochondrial membrane potential damage.

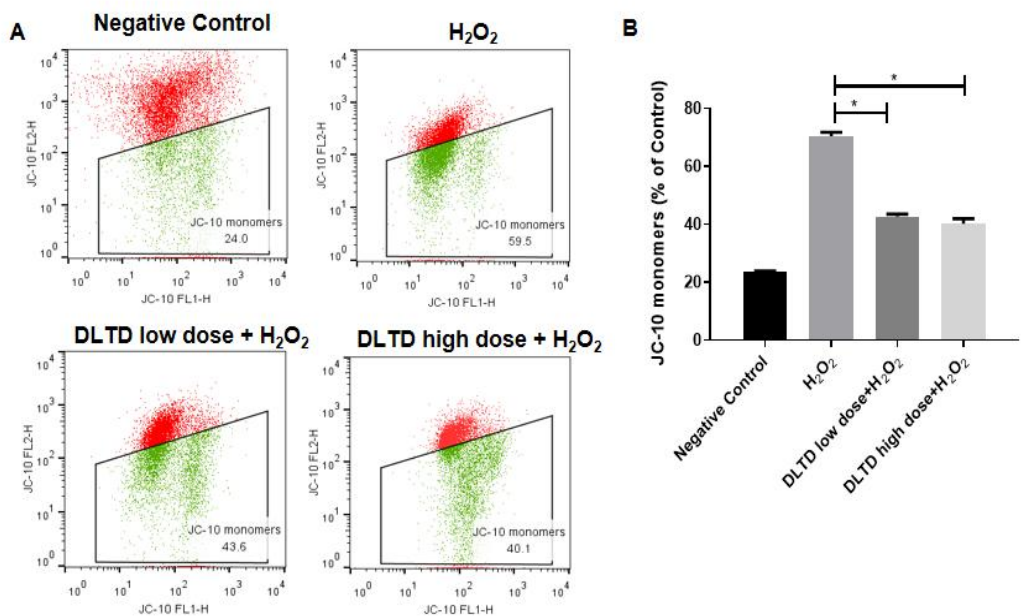

**Fig. 5.** DLTD inhibit the mitochondrial membrane potential damage of rat neural stem cell cells induced by hydrogen peroxide**.** A. Flow cytometry analysis the mitochondrial membrane potential damage of rat neural stem cell cells by JC-10 staining after different treatments with DLTD and $H_2O_2$. B. The fluorescence area of A (using the same threshold). Bar =100 μm.

**DLTD induced CDK2 and CDK4 and reduced CDKN1A protein expression level.** Through network pharmacology analysis, we found that DLTD targets the negative regulation of the apoptosis pathway gene CDKN1A. CDKN1A can encode p21 protein, this protein can bind to CDK2/4 and inhibit the activity of cyclin-CDK2 or cyclin-CDK4 complexes [16]. In order to detect whether DLTD reducing apoptosis level is through CDKN1A/CDK2/CDK4, we analyzed the expression of the three proteins in Rat neural stem cells with DLTD treatment. The ELISA results showed that the expression of CDK2 and CDK4 proteins of DLTD treatment group were much higher than those in $H_2O_2$ group, while the expression of CDKN1A proteins of DLTD treatment group were much lower than $H_2O_2$ group. And it was in a dose-dependent manner (Fig. 6).

In conclusion，DLTD could induce the expression of CDK2 and CDK4 proteins, and down regulate CDKN1A protein expression, which may contribute to the reduced ROS and apoptosis level.

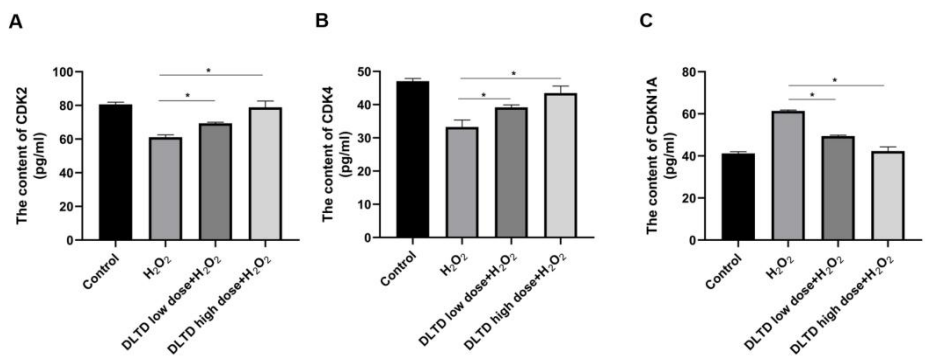

**Fig. 6.** ELISA analysis of the expression of three proteins after DLTD treatment. A. The expression level of CDK2 after DLTD treatment in Rat neural stem cells. B. The expression level of CDK4 after DLTD treatment in rat neural stem cells. C. The expression level of CDKN1A after DLTD treatment in Rat neural stem cells. Data are presented as the mean ± s.d. n = 3. *P < 0.05, **P < 0.01, ***P < 0.001.

For the treatment of diseases, TCM always take into consideration of complex symptoms and the whole-body situation [27,28]. LSS is a chronic disease with high surgical risk and may cause paralysis. DLTD has been used in the conservative treatment of LSS for decades. However, due to the complexity of TCM, the molecular mechanism of DLTD in treating LSS was not completely understood. In this study, we explored the potential mechanism of DLTD in the treatment of LSS. We found that DLTD treats LSS through reducing ROS and apoptosis levels by inhibiting CDKN1A and inducing CDK2 and CDK4 expression levels.

A plenty of active components involved in DLTD have been proved to have anti-inflammatory, anti-apoptotic, and anti-oxidative stress effects. Astragaloside IV (AS-IV), one of the main active components of HuangQi, has been reported to effectively reduce apoptosis level in brain cells [29]. Furthermore, Aucubin is a natural compound in DuZhong. It was found that Aucubin may protect articular cartilage and slow the progression of Osteoarthritis (OA) by inhibiting chondrocyte apoptosis and excessive ROS production [30]. Xu et al. found that DanShen alleviates cartilage injury in rabbit OA and SNP induced apoptosis by inhibiting NF-κB signaling pathway [31].

Through network pharmacology analysis and the construction of DLTD-Targets-LSS disease network, we collected 23 intersecting genes. Among them, MYC, FOS, and CDKN1A are related to apoptosis and aging; NOS2, TNF, CALCR, IL10, IL1B, CXCL8 and IL6 are related to inflammation; and MMP1 and MMP9 are related to arthritis. Based on the previous studies, they have been used as marker genes to detect apoptosis, antioxidation, aging and inflammation [32-41]. Relevant studies have shown that MYC overexpression can induce ROS expression under some conditions [42]. In the future, we will continue to explore the interaction between them and further reveal the specific molecular mechanism of relieving LSS.

Among the 23 genes, we mainly focused on CDKN1A. CDKN1A is a negative regulation gene of cell cycle, which inhibits cell cycle progression. It is the target protein of two small molecules in DLTD, one is Tanshinone IIA (TSA) in DanShen, and the other is Quercetin, which is found in HuangQi, YanHuSuo and DuZhong. TSA is a phenanthraquinone derivative extracted from DanShen. It is one of the main active components of DanShen. In recent years, researchers found that TSA could inhibit apoptosis by decreasing the expression of cleaved Caspase-3 protein and increasing the expression of Bcl-2 protein [43,44]. Quercetin is a flavonoid, which has the functions of antioxidant, anticancer, anti-inflammatory, anti-aggregatory and anti-aging [45]. The study showed that quercetin can scavenge ROS and reduce the damage of oxidative stress [46]. Moreover, quercetin could reduce the response to inflammatory mediators and inhibit apoptosis [47]. Our work showed that these small molecules in the DLTD may directly or indirectly act on CDKN1A, thereby regulating the cell cycle and alleviating cell apoptosis，and then alleviating the symptoms of LSS. The interaction between quercetin and CDKN1A will be verified in the future work.

In conclusion, our study shows that DLTD is an effective drug to treat LSS through reducing ROS content, decreasing apoptosis by inhibiting CDKN1A and promoting CDK2 and CDK4 expression levels. It provides certain experimental data and evidence for the study of DLTD tablets in the treatment of LSS disease.

**Author contributions.** Jun Kang―conceived the project; Xue Bai and Wenbo Cheng―performed the experiments and analysis; Jun Kang and Ayesha T. Tahir―wrote the paper. All authors read and approved the final manuscript.

**Acknowledgments.** This work was financially supported by the National Science Foundation of China 3227100132. We thank Prof. Peiyuan Liu for his valuable suggestions and constructive comments on the manuscript.

**Conflicts of interest.** All authors declare no conflict of interest.